\documentclass[twocolumn,aps,showpacs]{revtex4}
\usepackage{amssymb}
\usepackage{epsf}

\usepackage{amsmath}

\def\lsim{\lower -0.3ex \hbox{$<$} \kern -0.75em \lower 0.7ex \hbox{$\sim$}}
\def\gsim{\lower -0.3ex \hbox{$>$} \kern -0.75em \lower 0.7ex \hbox{$\sim$}}

\def\Vec#1{{\bf #1}}

\def\vare{\varepsilon}

\begin{document}

\title{Trigonal warping and Berry's phase $N\pi$ in ABC-stacked
multilayer graphene}
\author{Mikito Koshino$^{1}$ and Edward McCann$^{2}$}
\affiliation{
$^{1}$Department of Physics, Tokyo Institute of
Technology, 2-12-1 Ookayama, Meguro-ku, Tokyo 152-8551, Japan\\
$^{2}$Department of Physics, Lancaster University, Lancaster, LA1
4YB, UK}

\begin{abstract}
The electronic band structure of ABC-stacked multilayer graphene
is studied within an effective mass approximation. The electron
and hole bands touching at zero energy support chiral
quasiparticles characterized by Berry's phase $N\pi$ for
$N$-layers, generalizing the low-energy band structure of
monolayer and bilayer graphene. We investigate the
trigonal-warping deformation of the energy bands and show that the
Lifshitz transition, in which the Fermi circle breaks up into
separate parts at low energy, reflects Berry's phase $N \pi$. It
is particularly prominent in trilayers, $N=3$, with the Fermi
circle breaking into three parts at a relatively large energy that
is related to next-nearest-layer coupling. For $N=3$, we study the
effects of electrostatic potentials which vary in the stacking
direction, and find that a perpendicular electric field, as well
as opening an energy gap, strongly enhances the trigonal-warping
effect. In magnetic fields, the $N=3$ Lifshitz transition is
manifested as a coalescence of Landau levels into
triply-degenerate levels.
\end{abstract}


\maketitle

\section{Introduction}\label{intro}

Soon after the fabrication of individual graphene flakes a few
years ago \cite{novo04}, it was realized that low-energy
quasiparticles in graphene are chiral, with a linear dispersion
and degree of chirality characterized by Berry's phase $\pi$ in
monolayer graphene \cite{novo05,zhang05} and quadratic dispersion
related to Berry's phase $2\pi$ in bilayers \cite{novo06,mcc06a}.
In addition to their degree of chirality, bilayers are
distinguished from monolayers by the possibility of using doping
or external gates to induce interlayer asymmetry that opens a
tunable gap between the conduction and valence bands
\cite{mcc06a,lu06,guinea06,mcc06b,min07,aoki07,gava09}, as
observed in transport \cite{castro07,oost08} and spectroscopic
measurements \cite{ohta06,li09,zhang08,kuzmenko,zhang09,mak09}.

Recently, there has been experimental interest in the transport
properties of trilayer graphene \cite{ohta07,guett08,crac08}. It
is expected that two different types of stacking order, ABA and
ABC (illustrated in Fig.~1), will be realized in nature and that
electronic properties will depend strongly on the stacking type.
For ABA-stacked trilayer graphene, the low-energy electronic band
structure consists of separate monolayer-like and bilayer-like
bands
\cite{lu06,latil06,part06,guinea06,Kosh_mlg,aoki07,kosh09,avet09}
that become hybridized in the presence of interlayer asymmetry
\cite{guinea06,kosh09}. By contrast, the low-energy bands of
ABC-stacked trilayers \cite{latil06,guinea06,lu06abc,aoki07} do not
resemble those of monolayers or bilayers, but appear to be a cubic
generalization of them. Thus, there is a cubic dispersion relation
and chirality related to Berry's phase $3 \pi$
\cite{guinea06,manes07,min08}, and, as in bilayers, the
application of interlayer asymmetry is predicted to open an energy
gap in the spectrum \cite{guinea06,aoki07}.

In this paper, we show that the low-energy band structure of
ABC-stacked multilayer graphene is not just a straightforward
generalization of that of monolayers and bilayers. We focus on a
particular aspect of the band structure, trigonal warping, which
plays a crucial role in the low-energy band structure. Trigonal
warping is a deformation of the Fermi circle around a degeneracy
point \cite{ando}, at each of two inequivalent corners of the
hexagonal Brillouin zone that are known as $K$ points
\cite{kpoints} [Fig.~1(b)]. In bilayer graphene, trigonal warping
is enhanced by the interlayer coupling and leads to a Lifshitz
transition \cite{lif} when the Fermi line about each $K$ point is
broken into several pockets
\cite{mcc06a,part06,Kosh_bilayer,cserti,manes07,mcc07,mik08,toke}.
Here, we develop an effective Hamiltonian for ABC-stacked trilayer
graphene, to show that trigonal warping in it is both
qualitatively and quantitatively different from that in bilayers.
The main contribution to trigonal warping arises from a different
type of interlayer coupling that is missing in bilayers and we
predict that it leads to a Lifshitz transition at a much larger
energy $\sim$ 10meV, which is 10 times as large as in a bilayer.
Moreover, on undergoing the Lifshitz transition, the Fermi surface
breaks into a different number of pockets reflecting Berry's phase
$3\pi$ in contrast to $2\pi$ in bilayers. Here, we also generalize
our approach to describe trigonal warping in general ABC-stacked
$N$-layer graphene, to show that Berry's phase $N\pi$ manifests
itself in different characteristics of the Lifshitz transition.

\begin{figure}[t]
\centerline{\epsfxsize=0.95\hsize \epsffile{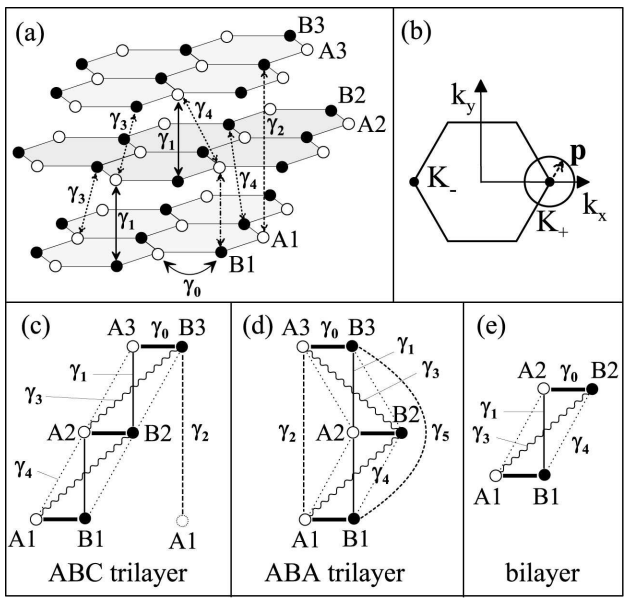}}
\caption{(a) Schematic of the ABC-stacked trilayer lattice
containing six sites in the unit cell, $A$ (white circles) and $B$
(black circles) on each layer, showing the
Slonczewski-Weiss-McClure parameterization \cite{dressel02} of
relevant couplings $\gamma_0$ to $\gamma_4$. (b) Schematic of the
hexagonal Brillouin zone with two inequivalent valleys $K_{\pm}$
showing the momentum $\mathbf{p}$ measured from the center of
valley $K_{+}$. Schematic of the unit cell of (c) ABC-stacked
trilayer graphene, (d) ABA-stacked trilayer graphene, and (e)
bilayer graphene. In (c), $\gamma_2$ describes a vertical coupling
between sites $B3$ and $A1$ in different unit cells.
}
\label{fig:1}
\end{figure}

In the next Section, we describe the effective mass model of
ABC-stacked trilayer graphene and the resulting band structure.
Then, in Section~\ref{effham}, we derive an effective low-energy
Hamiltonian and we use it to compare the behavior of low-energy
chiral quasiparticles in trilayers with those in monolayer and
bilayer graphene. In Section~\ref{liftrans}, we provide an
approximate analytical description of the Lifshitz transition in
the absence and in the presence of interlayer asymmetry that opens
a gap in the spectrum. Section~\ref{llspectrum} describes the
manifestation of the Lifshitz transition in the degeneracy of
Landau levels in the presence of a finite magnetic field. In
Section~\ref{general}, we generalize our approach to ABC-stacked
$N$-layer graphene. Throughout, we compare the approximate
description of the effective low-energy Hamiltonian with numerical
diagonalization of the full effective mass model.

\section{The effective mass model of ABC-stacked trilayer graphene}\label{model}

The lattice of ABC-stacked trilayer graphene consists of three
coupled layers, each with carbon atoms arranged on a honeycomb
lattice, including pairs of inequivalent sites $\{ A1 , B1 \}$,
$\{ A2 , B2 \}$, and $\{ A3 , B3 \}$ in the bottom, center, and
top layers, respectively. The layers are arranged as shown in
Fig.~\ref{fig:1}(a,c), such that pairs of sites $B1$ and $A2$, and
$B2$ and $A3$, lie directly above or below each other [for
comparison, the unit cell of ABA-stacked graphene is shown in
Fig.~\ref{fig:1}(d)]. In order to write down an effective mass
Hamiltonian, we adapt the Slonczewski-Weiss-McClure
parameterization of tight-binding couplings of bulk graphite
\cite{dressel02}. Nearest-neighbor ($Ai$-$Bi$ for $i= \{1,2,3\}$)
coupling within each layer is described by parameter $\gamma_0$,
$\gamma_1$ describes strong nearest-layer coupling between sites
($B1$-$A2$ and $B2$-$A3$) that lie directly above or below each
other, $\gamma_3$ ($\gamma_4$) describes weaker nearest-layer
coupling between sites $A1$-$B2$ and $A2$-$B3$ ($A1$-$A2$,
$B1$-$B2$, $A2$-$A3$, and $B2$-$B3$). With only these couplings,
there would be a degeneracy point at each of two inequivalent
corners, $K_{\pm}$, of the hexagonal Brillouin zone \cite{kpoints}
but this degeneracy is broken by next-nearest-layer coupling
$\gamma_2$, between sites $A1$ and $B3$ that lie on the same
vertical line \cite{latil06,lu06abc,aoki07}. For typical values of
bulk ABA graphite we quote \cite{dressel02} $\gamma_0 = 3.16$eV,
$\gamma_1 = 0.39$eV, $\gamma_2 = -0.020$eV, $\gamma_3 = 0.315$eV
and $\gamma_4 = 0.044$eV. Although the atomic structures of ABA
and ABC (rhombohedral) graphite are different, we refer to those
values in the following numerical calculations, assuming that the
corresponding coupling parameters have similar values
\cite{mcclure69}.

In a basis with atomic components $\psi_{A1}$, $\psi_{B1}$,
$\psi_{A2}$, $\psi_{B2}$, $\psi_{A3}$, $\psi_{B3}$, the
ABC-stacked trilayer Hamiltonian
\cite{mcclure69,guinea06,lu06abc,arovas08} is
\begin{eqnarray}
 \hat{H}_{ABC} =
\begin{pmatrix}
 D_1 & V  & W \\
 V^{\dagger} & D_2 & V  \\
 W^{\dagger} & V^{\dagger} & D_3
\end{pmatrix},
\label{h1}
\end{eqnarray}
where the $2 \times 2$ blocks are
\begin{eqnarray}
&& D_i =
\begin{pmatrix}
 U_i & v \pi^\dagger
 \\ v \pi & U_i
\end{pmatrix} \quad (i=1,2,3), \label{di}
\\
&&
V =
\begin{pmatrix}
-v_4 \pi^\dagger & v_3\pi \\ \gamma_1 & -v_4 \pi^\dagger
\end{pmatrix},
\quad
W =
\begin{pmatrix}
0 & \gamma_2/2 \\ 0 & 0
\end{pmatrix}, \label{vi}
\end{eqnarray}
%
%
where $v= ( \sqrt{3}/2 ) a\gamma_{0}/\hbar$, $v_3 = ( \sqrt{3}/2 )
a\gamma_{3}/\hbar$, $v_4 = ( \sqrt{3}/2 ) a\gamma_{4}/\hbar$, $\pi
= \xi p_x + i p_y$, $\pi^{\dag} = \xi p_x - i p_y$, and $\xi = \pm
1$ is the valley index. Here $\mathbf{p} = (p_x,p_y) = p (\cos
\phi , \sin \phi )$ is the momentum measured with respect to the
center of the valley [Fig.~\ref{fig:1}(b)]. The parameters $U_1$,
$U_2$, and $U_3$ describe on-site energies of the atoms on the
three layers that may be different owing the presence of
substrates, doping, or external gates. In the following, we set
the average on-site energy to zero $U_1 + U_2 + U_3 = 0$ and write
differences between the on-site energies in terms of asymmetry
parameters $\Delta_1$ and $\Delta_2$ \cite{kosh09},
\begin{eqnarray*}
\Delta_1 &=& \left( U_1 - U_3 \right)/2 \, , \\
\Delta_2 &=& \left( U_1 - 2U_2 + U_3 \right)/6 \, .
\end{eqnarray*}
Parameter $\Delta_1$ describes a possible asymmetry between the energies
of the outer layers, whereas $\Delta_2$ takes into account the possibility
that the energy of the central layer may differ from the average outer layer
energy.

As there are six atoms in the unit cell, ABC-stacked trilayer
graphene has six electronic bands at low energy as plotted in
Fig.~2. For no interlayer asymmetry $\Delta_1 = \Delta_2 = 0$, and
exactly at the K point, $p = 0$, the eigenvalues $\epsilon$ of
Hamiltonian Eq.~(\ref{h1}) are given by $(\epsilon^2 - \gamma_1^2
)^2 (\epsilon^2 - \gamma_2^2/4) = 0$. Four of the bands are split
away from zero energy by interlayer coupling $\gamma_1$ ($\epsilon
= \pm \gamma_1$ twice). These high-energy bands correspond to
dimer states formed primarily from orbitals on the atomic sites
$B1$, $A2$, $B2$, and $A3$ that are strongly coupled by
$\gamma_1$. The other two bands ($\epsilon = \pm \gamma_2/2$) are
split slightly away from zero energy by next-nearest layer coupling
$\gamma_2/2$ that connects atomic sites $A1$ and $B3$
\cite{latil06,lu06abc,aoki07}.

\begin{figure}[t]
\centerline{\epsfxsize=0.7\hsize \epsffile{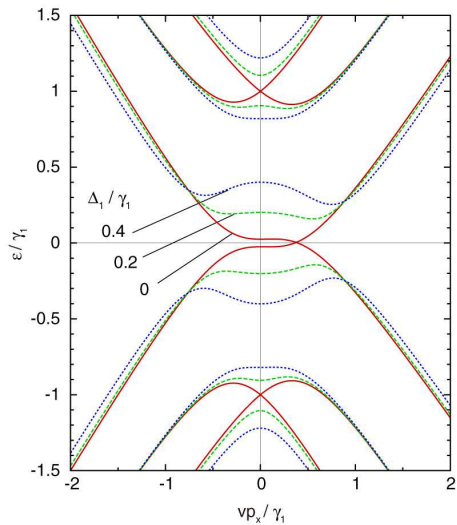}}
\caption{Band dispersion of ABC-stacked trilayer graphene in the
vicinity of $K_{+}$ along $p_x$ axis. Parameter values are
$\gamma_0 = 3.16$eV, $\gamma_1 = 0.39$eV, $\gamma_2 = -0.020$eV,
$\gamma_3 = 0.315$eV and $\gamma_4 = 0.044$eV \cite{dressel02}.}
\label{fig:band}
\end{figure}

\begin{figure}[t]
\centerline{\epsfxsize=0.7\hsize \epsffile{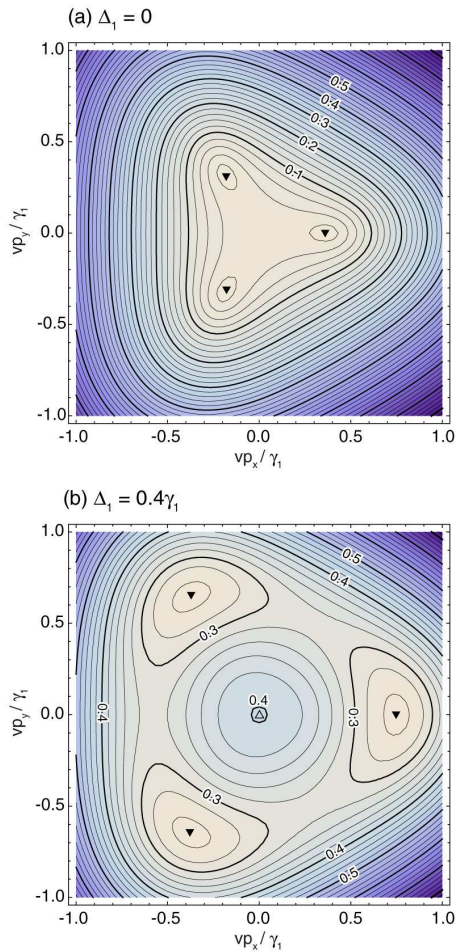}}
\caption{(a) Equi-energy contour plots of the lowest electron band
of ABC trilayer graphene at (a) $\Delta_1 = 0$ and (b)
$0.4\gamma_1$. Numbers on the contours indicate
energy in units of $\gamma_1$.
Filled and empty triangles represent local minima
and maxima, respectively, of the energy band.}
\label{fig:contour}
\end{figure}

Figure \ref{fig:band} shows the band structure at several
$\Delta_1$'s with $\Delta_2 =0$, using the parameter values quoted
above. $\Delta_1$ opens an energy gap between the lower electron
and hole bands, because of the energy difference between $A1$ and
$B3$ sites \cite{guinea06,aoki07}. Figure \ref{fig:contour} shows
contour plots of the lower electron band at (a) $\Delta_1/\gamma_1
= 0$ and (b) 0.4, showing that the band is trigonally warped, and
the contour splits into three pockets at low energy. The detailed
band structure and its relation to the band parameters will be
studied in the following sections.

\section{The low-energy effective Hamiltonian}\label{effham}

To describe the low-energy electronic properties of ABC-stacked
trilayer graphene it is useful to derive an effective
two-component Hamiltonian that describes hopping between atomic
sites $A1$ and $B3$. Such a procedure has been applied to bilayer
graphene \cite{mcc06a} and to ABC-stacked trilayer graphene
\cite{guinea06} for $\gamma_3 = \gamma_2 = \Delta_2 = 0$. We begin
with the energy eigenvalue equation $H \Psi = \epsilon \Psi$ of
the six-component Hamiltonian, Eq.~(\ref{h1}), eliminate the
dimer components $\chi = \left( \psi_{B1}, \psi_{A2}, \psi_{B2},
\psi_{A3} \right)^{T}$ and, then, simplify the expressions for
low-energy components $\theta = \left( \psi_{A1}, \psi_{B3}
\right)^{T}$ by treating interlayer coupling $\gamma_1$ as a large
energy scale $|\epsilon |, vp, |\gamma_2 |, |\gamma_3 |, |\gamma_4
|, |\Delta_1 |, |\Delta_{2}| \ll \gamma_{1}$.
We denote $h_{\theta}$
as the diagonal block of Hamiltonian
of Eq. (\ref{h1}) corresponding to $\theta$,
$h_{\chi}$ as the four by four diagonal block corresponding to $\chi$,
and $u$ as the off-diagonal $2\times 4$ block
coupling $\theta$ and $\chi$.
The Schr\"odinger equation for $\theta$
can be expanded up to first order in $\epsilon$ as
$[h_\theta - u h_\chi^{-1}u^\dagger]\theta = \epsilon S \theta$
with $S\equiv 1 + u h_\chi^{-2} u^\dagger$.
Then, the effective Hamiltonian for $\tilde\theta = S^{1/2}\theta$ becomes
${H}^{\rm (eff)} \approx S^{-1/2} [h_\theta - u h_\chi^{-1}u^\dagger] S^{-1/2}$.

Thus, we find the following two-component Hamiltonian in a
basis of the $A1$-$B3$ sites:
\begin{eqnarray}
{\hat{H}}^{\rm (eff)}_{ABC}
&=& {\hat{H}}_{3} + {\hat{H}}_{3w} + {\hat{H}}_{3c}  +{\hat{H}}_{\Delta 1} +
{\hat{H}}_{\Delta 2}  ,  \label{heff1} \\
{\hat{H}}_{3} &=& \frac{v^3}{\gamma_1^2}\left(
\begin{array}{cc}
0 & \left( {\pi }^{\dag }\right) ^{3} \\
{\pi ^{3}} & 0
\end{array} \right) , \nonumber \\
{\hat{H}}_{3w} &=&
\left(
- \frac{2 v v_3 p^2}{\gamma_1} + \frac{\gamma_2}{2}
\right)
 \left(
\begin{array}{cc}
0 & 1 \\
1 & 0
\end{array}
\right) , \nonumber
\\
{\hat{H}}_{3c} &=&
\frac{2 v v_4 p^2}{\gamma_1^2}
\left(
\begin{array}{cc}
1 & 0 \\
0 & 1
\end{array}
\right). \nonumber
\\
{\hat{H}}_{\Delta 1} &=& \Delta_1
\left(
1-\frac{v^2p^2}{\gamma_1^2}
\right)
\left(
\begin{array}{cc}
1 & 0 \\
0 & - 1
\end{array}
\right)
,  \nonumber
\\
{\hat{H}}_{\Delta 2} &=&
\Delta_2
\left(
1- \frac{3 v^2p^2}{\gamma_1^2}
\right)
\left(
\begin{array}{cc}
1 & 0 \\
0 & 1
\end{array}
\right)
.  \nonumber
\end{eqnarray}
Here we keep only the leading order for the terms including
$\gamma_2$, $v_3$ and $v_4$. Terms ${\hat{H}}_{3}$ and
${\hat{H}}_{\Delta 1}$ were derived in Ref.~\cite{guinea06}. The
cubic term ${\hat{H}}_{3}$ describes effective hopping between
sites $A1$ and $B3$ via the other sites on the lattice that are
strongly coupled by $\gamma_1$. Taken on its own, it produces a
dispersion $\epsilon = \pm v^3 p^3 / \gamma_1^2$. ${\hat{H}}_{3w}$
arises from the skewed interlayer coupling $\gamma_3$ and the
next-nearest interlayer coupling $\gamma_2$, and is responsible
for trigonal warping as discussed in detail later.
${\hat{H}}_{3c}$, coming from another interlayer coupling
$\gamma_4$, gives an identical curvature to the electron and hole
bands and thus introduces electron-hole asymmetry. Terms
${\hat{H}}_{\Delta 1}$, ${\hat{H}}_{\Delta 2}$ arise from the
interlayer asymmetries $\Delta_1$ and $\Delta_2$, respectively.
${\hat{H}}_{\Delta 1}$ leads to the opening of an energy gap
between the conduction and valence bands, while ${\hat{H}}_{\Delta
2}$ produces electron-hole asymmetry in a similar way as
${\hat{H}}_{3c}$. In the two-component basis of ${\hat{H}}^{\rm
(eff)}_{ABC}$, time reversal is described by ${\hat{H}}^{\ast}
\left({\mathbf{p},\Delta_1,\xi}\right) = {\hat{H}}
\left(-{\mathbf{p},\Delta_1,-\xi}\right)$ and spatial inversion by
$\sigma_x {\hat{H}} \left({\mathbf{p},\Delta_1,\xi}\right)
\sigma_x = {\hat{H}} \left(-{\mathbf{p},-\Delta_1,-\xi}\right)$.
Manes {\em et al} \cite{manes07} showed that the Fermi points of
ABC-stacked multilayers are stable with respect to the opening of
a gap against perturbations that respect combined time reversal
and spatial inversion, as well as translation invariance.

The low-energy effective Hamiltonian for ABC-stacked trilayer
graphene bears some resemblance to that of bilayer graphene
\cite{mcc06a}. In the lattice of bilayer graphene, Fig.~1(e), two
of the sites ($B1$ and $A2$) are directly above or below each
other and are strongly coupled by interlayer coupling $\gamma_1$
whereas two sites ($A1$ and $B2$) do not have a counterpart in the
other layer. The low-energy Hamiltonian is written in a basis $(
\psi_{A1}, \psi_{B2})$ of these two sites:
\begin{eqnarray}
{\hat{H}}^{\rm (eff)}_{AB}
&=& {\hat{H}}_{2} + {\hat{H}}_{2w} + {\hat{H}}_{\Delta} ,  \label{heff2} \\
{\hat{H}}_{2} &=&-\frac{v^2}{\gamma_1}\left(
\begin{array}{cc}
0 & \left( {\pi }^{\dag }\right) ^{2} \\
{\pi ^{2}} & 0
\end{array} \right) , \nonumber \\
{\hat{H}}_{2w} &=& v_{3}\left(
\begin{array}{cc}
0 & {\pi } \\
{\pi }^{\dag } & 0
\end{array}
\right) , \nonumber
\\
{\hat{H}}_{2c} &=&
\frac{2 v v_4 p^2}{\gamma_1^2}
\left(
\begin{array}{cc}
1 & 0 \\
0 & 1
\end{array}
\right), \nonumber
\\
{\hat{H}}_{\Delta} &=&
\Delta
\left(
1- \frac{2 v^2p^2}{\gamma_1^2}
\right)
\left(
\begin{array}{cc}
1 & 0 \\
0 & -1
\end{array}
\right), \nonumber
\end{eqnarray}
where parameter $\Delta$ describes interlayer asymmetry between
on-site energy $\Delta$ of the atoms, $A1$ and $B1$, on the first
layer and $-\Delta$ of the atoms, $A2$ and $B2$, on the second
layer.

The first term in each Hamiltonian, ${\hat{H}}_{2}$ for bilayers,
Eq.~(\ref{heff2}), and ${\hat{H}}_{3}$ for ABC-stacked trilayers,
Eq.~(\ref{heff1}), are members of a family of Hamiltonians
${\hat{H}}_{J}= F(p)\,{\boldsymbol\sigma }\cdot \mathbf{n}$ where
$\mathbf{n} = \mathbf{l}_{x} \cos (J\xi\phi) + \mathbf{l}_{y} \sin
(J\xi\phi )$ for $\mathbf{p} = p(\cos \phi , \sin \phi )$
\cite{mcc06a,guinea06,manes07,min08}. They describe chiral
quasiparticles, and the degree of chirality is $J=1$ in monolayer
graphene, $J=2$ in a bilayer, and, here, $J=3$ in ABC-stacked
trilayer. Quasiparticles described by the Hamiltonians
${\hat{H}}_{J}$ acquire a Berry's phase $ -i \oint_C d\Vec{p}\cdot
\langle \Psi|\nabla_{\Vec{p}}|\Psi \rangle = J\xi\pi$, upon an
adiabatic propagation along an equi-energetic line $C$. Thus
charge carriers in ABC-stacked trilayer graphene are Berry's phase
$3\xi \pi$ quasiparticles, in contrast to Berry's phase $\xi\pi$
particles in monolayers, $2\xi\pi$ in bilayers. As well as the
first term in the Hamiltonian Eq.~(\ref{heff1}) of ABC-trilayers
being a generalization of that in bilayers, the influence of
interlayer asymmetry $\Delta_1 = ( U_1 - U_3 )/2$ as described by
${\hat{H}}_{\Delta 1}$ is similar to that in bilayers as described
by ${\hat{H}}_{\Delta}$, Eq.~(\ref{heff2}).

\section{Trigonal warping and the Lifshitz
transition}\label{liftrans}


In a similar way to bulk graphite \cite{dressel02}, the parameter
$\gamma_3$ [where $v_3 = ( \sqrt{3}/2 ) a\gamma_{3}/\hbar$]
produces trigonal warping in bilayer graphene \cite{mcc06a}, where
the equi-energetic line around each valley is stretched in three
directions. This is due to the interference of the matrix elements
connecting $A1$ and $B2$, where an electron hopping from $A1$ to
$B2$ acquires a factor $e^{2i\xi\phi}$ in $\hat{H}_2$ and
$e^{-i\xi\phi}$ in $\hat{H}_{2w}$. We neglect the terms including
$v_4$ which add a term $\propto p^2$ to the energy but don't
contribute to trigonal warping. At $\Delta =0$, the eigenenergy of
Eq.~(\ref{heff2}) is given by
\begin{eqnarray}
\epsilon \approx \pm \sqrt{v_3^2 p^2
- 2 \xi \frac{v_3v^2p^3}{\gamma_1}\cos 3\phi +
\frac{v^4p^4}{\gamma_1^2}} \, .
\end{eqnarray}
The warping has a dramatic effect when $\hat{H}_{2}$ and
$\hat{H}_{2w}$ have comparable amplitudes, i.e., $v^2p^2/\gamma_1
\sim v_3p$, which is satisfied at $p \sim p_0 = \gamma_1v_3/v^2$.
It leads to a Lifshitz transition
\cite{lif,mcc06a,part06,Kosh_bilayer,cserti,manes07,mcc07,mik08,toke},
in which the equi-energetic line is broken into four separate
pockets. There is one central pocket located around $p = 0$ and,
three \textquotedblleft leg\textquotedblright\ pockets centered at
momentum of magnitude $p=p_0$ and angle $\phi_{0} = 2n\pi/3 +
(1-\xi)\pi/6$. The Fermi pocket separation occurs at energy
$\epsilon_L = (v_3 / v)^2 \gamma_1 /4$, which is estimated to be
$\epsilon_L \sim 1$meV.


In ABC-stacked trilayer graphene, there is a similar, but much
greater warping effect. In hopping from $A1$ to $B3$, an electron
acquires a factor $e^{3i\xi\phi}$ from  $\hat{H}_3$ and a factor
of unity from $\hat{H}_{3w}$, giving trigonal symmetry in $\phi$.
At $\Delta_1 = \Delta_2 = 0$, the eigenenergy of
Eq.~(\ref{heff1}) reads,
\begin{eqnarray}
\epsilon \approx \pm \sqrt{f(p)^2
+ 2 \xi f \!\left( p  \right) g(p)\cos 3\phi +
g \!\left( p  \right)^2} \, ,
\label{eq_trig}
\end{eqnarray}
where $f(p) = v^3p^3/\gamma_1^2$ comes from  $\hat{H}_{3}$ and
$g(p) = -2v v_3 p^2/\gamma_1 + \gamma_2/2$ from $\hat{H}_{3w}$.
Similarly to the bilayer, the warping effect is prominent when
$|g(p)| \sim f(p)$, or $p \sim p_0$ with $vp_0/\gamma_1 \equiv
[\gamma_2/(2\gamma_1)]^{1/3} - (v_3/v)/3$. This estimate is valid
as long as  $|v_3/v| \lsim |\gamma_2/\gamma_1|^{1/3}$, which holds
for typical parameter values of bulk graphite \cite{dressel02}.

The major difference from the bilayer is the contribution of the
parameter $\gamma_2/2$, which appears in the Hamiltonian without
an accompanying momentum-dependent factor and, thus, it doesn't
vanish at $p=0$. Such trigonal warping produces a Lifshitz
transition at low energy, but, unlike bilayers, it occurs at
energy $\epsilon_L \approx |\gamma_2/2|$. Although the value of
$\gamma_2$ in ABC-trilayer graphene has not been measured
experimentally, comparison with similar couplings in bulk graphite
\cite{dressel02} suggest that $|\gamma_2| \sim 20$meV. This opens
up the possibility that the Lifshitz transition in ABC-trilayer
graphene could occur at a much higher energy than that in
bilayers. At energy lower than $|\gamma_2|/2$, the contour splits
into three leg pockets centered at $p \sim p_0$ in a trigonal
manner.
Unlike bilayer graphene, the central pocket is missing because
$\hat{H}_{\gamma_2}$ does not vanish at $p=0$.

An effective Hamiltonian in the vicinity of the leg pockets, for
$|\epsilon | \ll \epsilon_L$, may be obtained by transforming to
momentum ${\mathbf q} = (q_x , q_y)$ measured from their centers,
\begin{eqnarray}
q_x &=& p_x \cos \phi_0 + p_y \sin \phi_0 - p_0 \, , \label{qxt} \\
q_y &=& - p_x \sin \phi_0 + p_y \cos \phi_0 \, , \label{qyt}
\end{eqnarray}
and taking the limit of infinitely large $\gamma_1$:
\begin{eqnarray}
\!\!\!\!{\hat{H}}_{ABC}^{\rm leg} &=&
3 v \left| \frac{\gamma_2}{2\gamma_1} \right|^{2/3} \!\!\left(
\begin{array}{cc}
0 & \xi \alpha q_x - i q_y  \\
\xi \alpha q_x + i q_y & 0
\end{array}
\right) \! , \label{eq_H_leg}
\end{eqnarray}
where $\alpha = 1 + (4 v_3 / 3v) (2\gamma_1 / \gamma_2)^{1/3}$.
Thus, the pockets are elliptical with dispersion $\epsilon \approx
\pm 3 |\gamma_2 /(2\gamma_1)|^{2/3} v \sqrt{\alpha^2 q_x^2 +
q_y^2}$. The different nature of the Lifshitz transition in
bilayers and ABC-stacked trilayers is a manifestation of Berry's
phase. In trilayer graphene, the geometrical phase integrated
around the equi-energy line of each pocket is $\xi\pi$ as in a
monolayer, giving $3\xi\pi$ in total. This is different from
bilayers, where $3\xi\pi$ arises from three leg pockets and $-\xi\pi$
from the center pocket gives $2\xi\pi$ in total \cite{manes07,mik08}.


Interlayer asymmetry $\Delta_1$ opens a gap in the spectrum and
produces a Mexican hat feature in the low-energy dispersion. \cite{guinea06}
The eigenenergy corresponding to Eq.~(\ref{heff1}) is given by
\begin{eqnarray}
\epsilon \approx \pm \sqrt{f(p)^2
+ 2 \xi f \!\left( p  \right) g(p)\cos 3\phi +
g \!\left( p  \right)^2 + h(p)^2},
\end{eqnarray}
with an extra term as compared to Eq.~(\ref{eq_trig}), $h(p) =
\Delta_1(1-v^2p^2/\gamma_1^2)$, coming from
$\hat{H}_{\Delta_1}$. For no trigonal warping [$g(p) = 0$], 
it yields $\epsilon^2 = \Delta_1^2 ( 1 - v^2p^2 /
\gamma_1^2)^2 + v^6p^6/\gamma_1^4$. The energy is $\epsilon = \pm
\Delta_1$ at zero momentum, but there is a minima located
isotropically about the center of the valley at finite momentum $p
= p_1 \approx (2/3)^{1/4} \sqrt{|\Delta_1 \gamma_1|}/v$ (for
$|\Delta_1| \ll |\gamma_1|$) at which the energy is $\epsilon =
\epsilon_1 \approx \pm \Delta_1 \left( 1 - (2/3)^{3/2} | \Delta_1
/ \gamma_1| \right)$.

In the presence of trigonal warping, there is an interplay between
the Mexican hat feature and the Lifshitz transition. In the large
gap regime, such that $|g| \ll f, h$, the circular edge of the
band bottom is trigonally distorted by the perturbation of $g(p)$,
making three pockets on it. The bottom of the pockets moves to
momentum $p = p_1 + \delta p_1$ with $v\delta p_1/\gamma_1 \approx
(\sqrt{6}/8)[\gamma_2/(2\Delta_1)] - (5/6)(v_3/v) $, and energy
$\epsilon = \epsilon_1 - \delta \epsilon_1$ with $\epsilon_1 =
(2/3)^{3/4}\sqrt{\Delta_1/\gamma_1} \,\left|\gamma_2/2 -
\sqrt{8/3} (v_3/v)\Delta_1\right|$. The area of the pocket in
k-space becomes of the order of $p_1 \delta p_1$, and the depth in
energy is of order $\delta \epsilon_1$, both of which increase as
$\Delta_1$ increases. This significant enlargement of the trigonal
pockets, in the presence of finite $\Delta_1$, is illustrated in
Figure~\ref{fig:contour} which is produced by numerical
diagonalization of the full Hamiltonian Eq.~(\ref{h1}). Note that
similar widening of the pockets by the gap term occurs in bilayer
graphene as well. This can be understood in an analogous way, by
writing $f(p) = v^2p^2/\gamma_1$, $g(p) = v_3 p$, and $h(p) =
\Delta(1-2v^2p^2/\gamma_1^2)$.

\section{Landau level spectrum}\label{llspectrum}

The energy levels in a magnetic field are given by replacing
$\Vec{p}$ with $\Vec{p} + e\Vec{A}$ in the Hamiltonian
Eq.~(\ref{h1}), where $\Vec{A}(\Vec{r})$ is the vector potential
corresponding to the magnetic field. Here we consider a uniform
magnetic field $B$ applied along $+z$ direction in a Landau gauge
$\Vec{A} = (0,Bx)$. Operators $\pi$ and $\pi^\dagger$ are then
related to raising and lowering operators $a^\dagger$ and $a$ of
the Landau level in a conventional two-dimensional system, such
that $[l_B/(\sqrt{2}\hbar)] \pi^\dagger =  a^\dagger$ and $a$ for
$K_+$ and $K_-$, respectively, with $l_B = \sqrt{\hbar/(eB)}$. The
operator $a$ acts as $a \varphi_{n,k} = \sqrt{n}\varphi_{n-1,k}$,
and $a \varphi_0 = 0$, where
$\varphi_{n,k}(x,y) \propto e^{iky} e^{-z^2/2}H_n(z)$ is the
wavefunction of the $n$th Landau level in a conventional
two-dimensional system with $z = (x+kl_B^2)/l_B$, and $H_n$ being
a Hermite polynomial.

In the simplest model including only $\gamma_0$
and $\gamma_1$ without trigonal warping, the effective Hamiltonian
$\hat{H}_3$ in Eq. (\ref{heff1}) yields the eigenstates for $K_+$
\cite{guinea06}
\begin{eqnarray}
&& \epsilon_{n} = 0, \quad
\Psi_{nk} \propto
 \begin{pmatrix}
  \varphi_{n,k} \\ 0
 \end{pmatrix} \quad (n=0,1,2),
\nonumber\\
&&
\left.
\begin{array}{l}
 \epsilon_{sn} = s \displaystyle\frac{\Delta_B^3}{\gamma_1^2}
\sqrt{n(n-1)(n-2)} \\
 \Psi_{snk} \propto
 \begin{pmatrix}
  \varphi_{n,k} \\ s \varphi_{n-3,k}
 \end{pmatrix}
\end{array}
\right\}
\quad (n\geq 3),
\label{eq_LL}
\end{eqnarray}
where $s=\pm 1$
describes the electron and hole levels, respectively, $\Delta_B =
\sqrt{2\hbar v^2 eB}$.
The eigenstates $n=0,1,2$ have a non-zero amplitude only
on the first element $(A1)$, and remain at zero energy regardless
of the magnetic field strength, while the energy of the other levels
behaves as $\propto B^{3/2}$. At the other valley $K_-$, there is
a similar structure except that the first and second elements are
interchanged, i.e., the zero-energy Landau levels have amplitudes
only on sites $B3$ \cite{guinea06}.

Trigonal warping gives a remarkable feature in the structure of
Landau levels. In enough small fields, the three leg pockets
independently accommodate an equal number of Landau levels so that
they are triply degenerate. This is in contrast to bilayer
graphene where the central pocket also contributes to the
degeneracy \cite{mcc06a}. The low-energy effective Hamiltonian,
Eq.~(\ref{eq_H_leg}), shows that the Landau level energy follows a
similar sequence as that in monolayer graphene,
$\epsilon_{n} =
3|\gamma_2/(2\gamma_1)|^{2/3} \sqrt{\alpha}\Delta_B {\rm sgn}(n)
\sqrt{n}$ where $n$ is integer. The total number of Landau levels
accommodated in each pocket is roughly estimated by the condition
$\vare_n \sim |\gamma_2|/2$, as $n \sim (\gamma_1/\Delta_B)^2
[\gamma_2/(2\gamma_1)]^{2/3} /(9\alpha)$.

\begin{figure}[t]
\centerline{\epsfxsize=0.7\hsize \epsffile{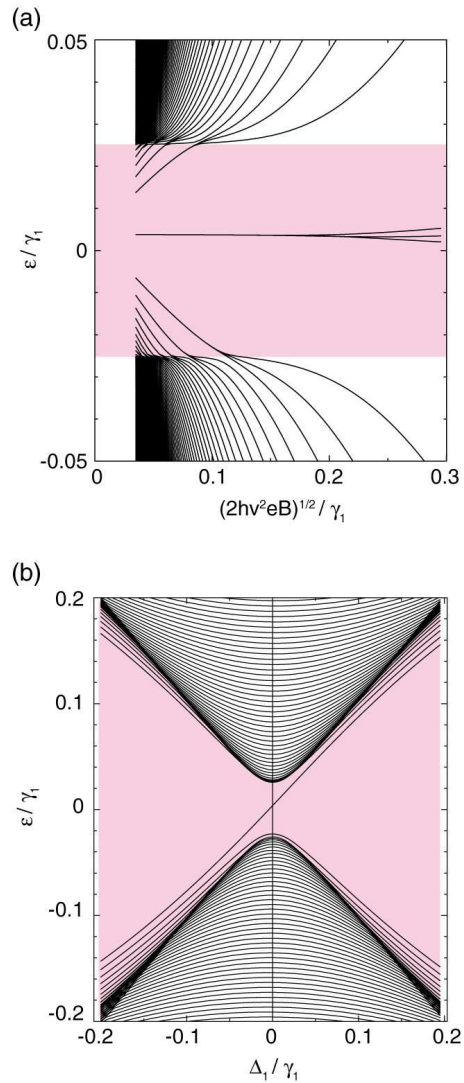}}
\caption{Landau levels of ABC trilayer graphene, plotted against
(a) $B^{1/2}$ at fixed $\Delta_1 = 0$, and (b) $\Delta_1$ at fixed
magnetic field $(2\hbar v^2 eB)^{1/2} =0.1\gamma_1 $ ($B\sim1$T). The region in
which Landau levels are triply degenerate is highlighted by
shading.} \label{fig:landau}
\end{figure}

Fig.~\ref{fig:landau}(a) shows the Landau level spectrum at the
valley $K_+$ as a function of $\Delta_B (\propto \sqrt{B})$,
numerically calculated for the full parameter model Eq. (\ref{h1})
at $\Delta_1=\Delta_2=0$. Below $\epsilon = \gamma_2/2$, the
Landau levels are triply degenerate and move in proportion to
$\sqrt{B}$. The degeneracy of each level is broken at $\epsilon =
\gamma_2/2$, and it splits into three separate levels,
corresponding to coalescence of the leg pockets at the Lifshitz
transition. At even higher energy, it approaches $B^{3/2}$
behavior as described in Eq.~(\ref{eq_LL}). The triply degenerate
level around zero energy is regarded as the $n=0$ level in each of
three pockets. In actual fact, its degeneracy is split slightly in
a large magnetic field, owing to magnetic break down among the
semi-classical orbits in the leg pockets, which is caused by the
parameter $v_4$.
When the trigonal warping vanishes,
those three levels switch to the degenerate levels
with indices $n=0,1,2$ in Eq.~(\ref{eq_LL}).

Fig.~\ref{fig:landau}(b) shows the Landau level spectrum at $K_+$
as a function of asymmetry $\Delta_1$ with fixed magnetic field
$\Delta_B =0.1\gamma_1 $ ($B\sim1$T).
As $\Delta_1$ is changed from negative to positive, three Landau
levels [indicated by the single diagonal line that crosses
$\epsilon = 0$ at $\Delta_1 = 0$ in Fig.~\ref{fig:landau}(b)] are
pumped from the hole side to the electron side. In the approximate
model of Eq.~(\ref{eq_LL}), this corresponds to the fact that the
energy levels $n=0,1,2$ have a wave amplitude only on $A1$, so
that it acquires on-site energy $+\Delta_1$ in the first order of
perturbation. At the other valley $K_-$, there is the opposite
movement, i.e., the three levels go down from positive to negative
energies in increasing $\Delta_1$.

The energy of the Lifshitz transition appears as a region where
the levels are densely populated, and below that energy the levels
are triply degenerated [indicated by the shaded region in
Fig.~\ref{fig:landau}(b)]. It should be noted that the number of
triply-degenerate levels increases for larger $\Delta_1$,
reflecting the enlargement of the trigonal pockets discussed
above. In a measurement of Hall conductivity, those
triply-degenerate Landau levels would be observed as quantum Hall
steps of magnitude $3g_vg_se^2/h$, where $g_v=g_s=2$ are the
valley and spin degeneracies, respectively.

\section{General ABC-stacked multilayer graphene}\label{general}

The analysis of ABC-stacked trilayer graphene can be extended to
multilayers with $N$ layers. We consider each layer to consist of
carbon atoms on a honeycomb lattice, and the layers are arranged
with ABC stacking. The Hamiltonian is written in a basis
$\psi_{A1}$, $\psi_{B1}$, $\psi_{A2}$, $\psi_{B2}$, $\cdots$,
$\psi_{AN}$, $\psi_{BN}$, as \cite{mcclure69,arovas08}
\begin{eqnarray}
 \hat H_{N} =
\begin{pmatrix}
 D_1 & V & W \\
 V^{\dagger} & D_2 & V & W \\
W^\dagger & V^\dagger & D_3 & \ddots & \ddots \\
& W^\dagger & \ddots & \ddots &   \\
& & \ddots
\end{pmatrix},
\label{eq_H_N}
\end{eqnarray}
where the $2 \times 2$ blocks $D_i$, $V$, $W$ are defined in
Eqs.~(\ref{di},\ref{vi}). Pairs of sites $B{(i)}$ and $A{(i+1)}$
$(i=1,\cdots,N-1)$ are vertically above or below each other, and
are strongly coupled by $\gamma_1$ giving dimer states. Thus, all
the sites in the lattice, except two, contribute to bands that lie
away from zero energy. The remaining two sites, $A_1$ and $B_N$,
form the lowest-energy electron and hole bands. Note that these
sites lie on the outer layers, so that the lowest bands are
missing in an infinite system with periodic boundary conditions
applied in the stacking direction. The band structure has trigonal
symmetry for any $N$. This is checked by applying the
transformation $\phi \rightarrow \phi+2\pi/3$ to
Eq.~(\ref{eq_H_N}), where the change in the matrix elements can be
canceled by the gauge transformation $\tilde\psi_{An} =  \alpha_n
\psi_{An}$ and $\tilde\psi_{Bn} = \alpha_n \psi_{Bn}$, with
$\alpha_n = e^{i\xi 2n\pi/3}$.

The effective low-energy Hamiltonian is obtained by treating terms
other than $\gamma_1$ as perturbations.
The effective Hamiltonian in a basis
$\{\psi_{A1}$, $\psi_{BN}\}$ reads
\begin{eqnarray}
&& \hat H^{\rm (eff)}_{N} =
\begin{pmatrix}
0 & X(p) \\
X^\dagger(p) & 0
\end{pmatrix}
+
\frac{2v v_4 p^2}{\gamma_1^2}
\begin{pmatrix}
1 & 0 \\
0 & 1
\end{pmatrix},
\nonumber \\
&& X(p) = \sum_{\{n_1,n_2,n_3 \}}
\frac{(n_1+n_2+n_3)!}{n_1!n_2!n_3!}
\frac{1}{(-\gamma_1)^{n_1+n_2+n_3-1}} \times \nonumber\\
&& \qquad\qquad\qquad (vp e^{i\xi\phi})^{n_1} (v_3p
e^{-i\xi\phi})^{n_2} \left(\frac{\gamma_2}{2}\right)^{n_3},
\label{eq_H_N_eff}
\end{eqnarray}
where the summation is taken over positive integers which satisfy
$n_1+2n_2+3n_3 = N$. Here we collected all the higher order terms
not including $v_4$, but retain just the leading term for $v_4$.
The trigonal warping structure can be described well in this
treatment as shown below, since $v_4$ only gives the
circularly-symmetric band curvature as in ABC trilayer.

\begin{figure}
\centerline{\epsfxsize=0.9\hsize \epsffile{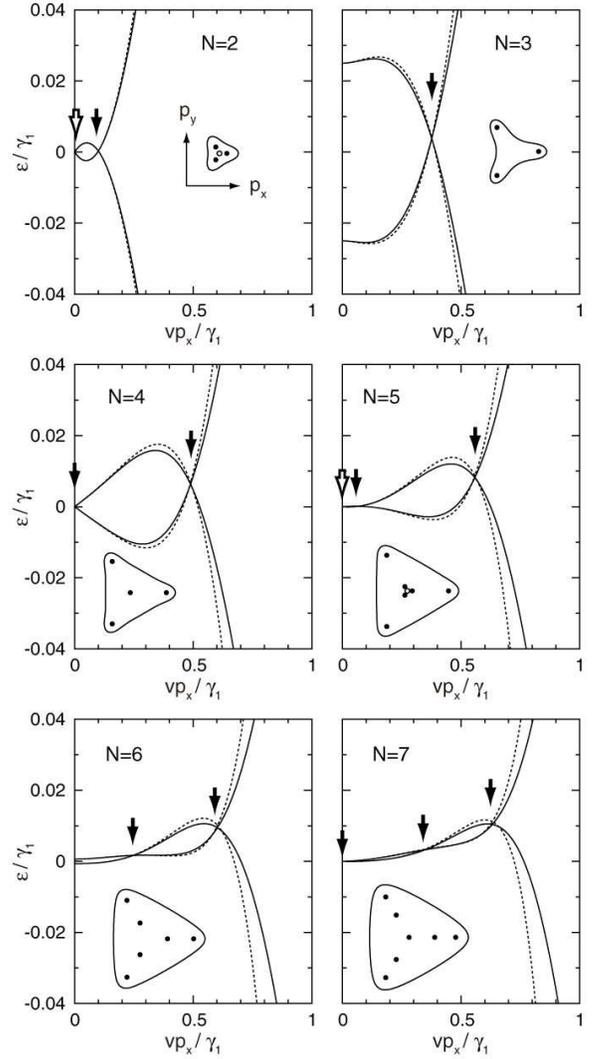}}
\caption{Low-energy band structure of ABC-stacked multilayer
graphene for several different layer numbers $N$, at $\Delta_1 =
\Delta_2 = 0$. Solid and dashed curves are calculated using
Eq.~(\ref{eq_H_N}) and its approximation Eq.~(\ref{eq_H_N_eff}),
respectively. Insets show the equi-energy lines at $\epsilon =
0.04\gamma_1$. The black and white arrows (circles in insets)
represent Dirac points having Berry's phase $\xi\pi$ and
$-\xi\pi$, respectively. } \label{fig:general}
\end{figure}

The eigenenergies are given by $\vare = 2vv_4p^2/\gamma_1^2 \pm
|X(p)|$. If we neglect $\gamma_2$ and $v_3$, we have $X = (vp
e^{i\xi\phi})^{N}/(-\gamma_1)^{N-1}$ which gives a pair of bands,
isotropic in momentum, which touch at the origin
\cite{mcc06a,guinea06,manes07,min08}. Berry's phase integrated
along an energy contour is $N\xi\pi$ at every energy.
Perturbation by $\gamma_2$ and $v_3$ produces trigonal warping as
observed in the trilayer. Figure \ref{fig:general} shows the lower
energy band structure for $\phi=0$ at several $N$'s, where the
solid lines are calculated using the original Hamiltonian
Eq.~(\ref{eq_H_N}), and the dashed lines use
Eq.~(\ref{eq_H_N_eff}). We can see that the effective Hamiltonian
reproduces the original band structure rather well including the
positions of the band touching points, except that the magnitude
in energy tends to be overestimated around $vp \sim \gamma_1$
where the perturbative approach fails.

The band touching points, or Dirac points, are given by the
solution of $X(p) = 0$. They appear in a series of $p$'s at only
three angles $\phi_0 = 2n\pi/3 + (1-\xi)\pi/6$, and around which
the Hamiltonian has a chiral structure similar to monolayer
graphene. We empirically found that the arrangement of these
points obeys the following rules: We have $[(N+1)/3]$ Dirac points
at $p\neq 0$ at each of three angles, and each of them has Berry's
phase $\xi \pi$. Here $[x]$ represents the greatest integer which
does not exceed $x$. The Dirac point at the center $(p=0)$ only
appears when $N$ is not a multiple of 3, and its Berry's phase is
$\xi \pi$ and $-\xi \pi$ when $N\equiv 1$ and $ -1$ (mod 3)
respectively. The total Berry's phase summed over all Dirac points
is always $N\xi\pi$, the same as the value without trigonal
warping. The energy scale for fine structure around the Dirac
points becomes smaller as $N$ increases, because the matrix
elements connecting $A_1$ and $B_N$ become higher order in $p$ for
larger $N$. We see that $N=3$ has the most prominent structure,
where $\gamma_2$ directly connects $A_1$ and $B_N$. The parameter
$v_4$ never opens a gap at the Dirac points but gives an energy
shift by $2vv_4p^2/\gamma_1^2$ and associated band curvature,
leading to misalignment of the Dirac point energies as shown in
Fig.~\ref{fig:general}. The curvature is independent of $N$
because it is due to the second order process from $A_1$ or $B_N$
to the nearest-neighboring dimer state.

The approach applied to the Landau levels of the trilayer in
Sec.~\ref{llspectrum} can be extended to the $N$-layer case. In
the simplest model including only $\gamma_0$ and $\gamma_1$, the
Landau levels at $K_+$ read
\begin{eqnarray}
&& \epsilon_{n} = 0, \quad
\Psi_{nk} \propto
 \begin{pmatrix}
  \varphi_{n,k} \\ 0
 \end{pmatrix} \quad (n=0,1,\cdots,N-1),
\\
&&
\left.
\begin{array}{l}
 \epsilon_{sn} = s \displaystyle\frac{\Delta_B^N}{\gamma_1^{(N-1)}}
\sqrt{n(n-1)\cdots(n-N+1)} \\
 \Psi_{snk} \propto
 \begin{pmatrix}
  \varphi_{n,k} \\ s \varphi_{n-N,k}
 \end{pmatrix}
\end{array}
\right\}
\quad (n\geq N),\nonumber\\
\end{eqnarray}
with $s=\pm 1$. The first and second elements are again
interchanged at the other valley $K_-$. The zero-energy level is
now $N$-fold degenerate per valley and per spin
\cite{guinea06,manes07,min08}. In presence of trigonal warping and
$v_4$, however, this is expected to split in accordance with the
discrepancy between the energies of different Dirac points shown
in Fig.~\ref{fig:general} for $B=0$, while some levels keep
threefold degeneracy owing to trigonal symmetry as in the trilayer
case. It is possible that electronic interactions may create
exotic collective modes in such highly-degenerate Landau levels,
but we leave the discussion of this for future studies.

\section{Conclusions}

In ABC-stacked multilayer graphene with $N$ layers, two low-energy
bands in the vicinity of each valley are formed from two
electronic orbitals that lie on the bottom and top layers of the
system. Such bands support chiral quasiparticles corresponding to
Berry's phase $N \pi$ \cite{mcc06a,guinea06,manes07,min08}. The
interplay between different types of interlayer coupling produces
trigonal warping, in which the Fermi circle around each valley is
stretched in three directions. At very low energy, trigonal
warping leads to a Lifshitz transition \cite{lif} when the Fermi
circle breaks up into separate pockets, in such a way that the
total Berry's phase is conserved. We predict that the Lifshitz
transition is particularly prominent in trilayers, $N=3$, with the
Fermi circle breaking into three parts at a relatively large
energy that is related to next-nearest-layer coupling.


\section{Acknowledgments}

The authors thank T. Ando, V.~I.~Fal'ko, and H.~Schomerus for
discussions. This project has been funded by EPSRC First Grant
EP/E063519/1, the Royal Society, and the Daiwa Anglo-Japanese
Foundation, and by Grants-in-Aid for Scientific Research from the
Ministry of Education, Culture, Sports, Science and Technology,
Japan.



\begin{thebibliography}{99}
\bibitem{kpoints} Corners of the hexagonal Brillouin zone are located at
wave vector $\mathbf{K}_{\xi }=\xi ({\textstyle\frac{4}{3}}\pi
a^{-1},0)$, where $\xi =\pm 1$ and $a$ is the lattice constant.
\end{thebibliography}
\end{document}